# ELECTRIC-FIELD GRADIENT AT Cd IMPURITIES IN $In_2O_3$. A FLAPW STUDY.

L. A. Errico, M. Rentería, G. Fabricius, and G. N. Darriba
*Departamento de Física, Facultad de Ciencias Exactas,
Universidad Nacional de La Plata, C.C. 67, 1900 La Plata, Argentina*

We report an *ab initio* study of the electric-field gradient tensor (EFG) at Cd impurities located at both inequivalent cationic sites in the semiconductor $In_2O_3$. Calculations were performed with the FLAPW method, that allows us to treat the electronic structure of the doped system and the atomic relaxations introduced by the impurities in the host lattice in a fully self-consistent way. From our results for the EFG (in excellent agreement with the experiments), it is clear that the problem of the EFG at impurities in $In_2O_3$ cannot be described by the point-charge model and antishielding factors.

## Introduction

Perturbed-Angular Correlations (PAC) and other hyperfine interaction measurements are widely used experimental techniques that provide local information on the interaction of a probe-nucleus with the surrounding electronic charge distribution [1]. An interpretation of such measurements can lead to a detailed knowledge of structural, electronic and magnetic properties of a solid [2]. One of the measured quantities, the quadrupole coupling constant $\nu_Q$, is proportional to the mayor component ($V_{33}$) of the electric-field gradient tensor (EFG) and the nuclear quadrupole moment $Q$. Due to the $r^{-3}$ dependence of the EFG from the charge sources, the EFG "felt" by a nucleus reflects sensitively the non-spherical electronic charge distribution around the nucleus. Therefore, the EFG is one of the most important clues for the study of the electronic structure in solids and the nature of the chemical bonding.

Usually, the probe-atom is an impurity dopant, which is present in either trace or small amounts in the host crystal. For this reason, interpreting experimental results involves understanding chemical differences between the probe atom and the indigenous ion replaced by the impurity. For long time, approaches for a quantitative calculation of the EFG were often based on the point-charge model (PCM). Since such calculations do not account for any onsite polarization, Sternheimer antishielding factors were introduced, whereas chemical bonding was usually neglected. The value of such calculations was often quite limited. For an accurate calculation of the EFG, the electronic configuration of the host, perturbed by the presence of the impurity, has to be determined. This can be done in the frame of the density-functional theory. In this sense, in 1999 we began a systematic study using the Full-Potential Linearized-Augmented Plane Waves (FLAPW) method of the EFG in doped oxides [3-5].

Among the binary oxides, those that crystallize in the bixbyite structure were subject of several PAC investigations using $^{111}$Cd as probe [6]. Due to the high degree of ionicity of these oxides, it was assumed in the past that PCM was accurate enough to exactly predict the EFG in these compounds [7, 8]. We report in this work a FLAPW study of the bixbyite $In_2O_3$ doped with Cd. Our calculations predict that Cd induce relaxations in the $In_2O_3$ host. The excellent agreement between our results for $V_{33}$ and the asymmetry parameter $\eta$ and the experiments, demonstrate that this approach is suitable to calculate accurately structural and electronic properties of the system and to provide a deeper understanding of them.

## Method of calculation

$In_2O_3$ crystallize in the cubic bixbyite structure ($a=1.0117_1$ Å [9]). In this structure, the cations form a nearly cubic face-center lattice in which six out of the eight tetrahedral sites are occupied by oxygen atoms. The unit cell consists of eight such cubes, containing 32 In and 48 O ions. Two nonequivalent cation sites, called $C$ and $D$, both with 6-fold O coordination, characterize the structure. Site $D$ is axially symmetric and can be described as an In surrounded by six O atoms at the corners of a distorted cube, leaving two corners of the same diagonal free. In site $C$ the cube is more distorted and the O atoms leave free two corners on a face diagonal. The positions of all atoms are determined by four parameters: the coordinate $u$ fixed the position of the $C$-type In (In($C$)) ions, while the O positions are given by the parameters $x$, $y$ and $z$ [9].

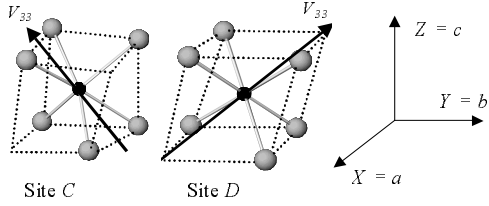

Figure 1: Scheme of the oxygen distribution around In atoms (black balls) in $In_2O_3$. In the real structure the oxygen atoms are slightly displaced from the corners of the cube. The arrows denotes the $V_{33}$ direction in each site. Results discussed in this work are referred to the indicated axes system (parallel to those defined by the crystalline axes).

In order to calculate from first-principles the EFG tensor at a Cd dilute impurity in $In_2O_3$ taking into account the structural and electronic effects introduced by the impurity in the host lattice, we considered the unit cell of $In_2O_3$ periodically repeated containing a single Cd-atom. Calculations were performed with the FLAPW [10] method as embodied in the WIEN97.10 code [11]. Exchange and correlation effects were treated using the generalized gradient (GGA) approximation [12]. In this method, the unit-cell is divided into non-overlapping spheres with radius $R_i$ and an interstitial region. The atomic spheres radii used for Cd, In and O were 1.11, 1.05 and 0.85 Å, respectively. We took for the parameter $RK_{MAX}$, which controls the size of the basis-set in these calculations, the value of 7. The correctness of the choice of these parameters was checked by performing calculations for $RK_{MAX}$=6 and 8. We also introduced local orbitals to include In-$4p$ and $4d$, O-$2s$ and Cd-$4p$ and $4d$ orbitals. Integration in reciprocal space was performed using the tetrahedron method taking 26 k-points (Cd at site $D$) and 60 k-points (Cd at site $C$) in the first Brillouin zone. Once self-consistency of the potential was achieved, the forces on the atoms were obtained [13], the ions were displaced [14] and the new positions for Cd neighbors were obtained. The procedure was repeated until the forces on the ions were below a tolerance value. At each step, the $V_{ii}$ elements of the EFG tensor were obtained from the $V_{2M}$ components of the lattice harmonic-expansion of the self-consistent potential [15].

There is still an important point to be taken into account: the charge state of the impurity system, the single acceptor $Cd^{2+}$ in $In_2^{3+}O_3^{2-}$. $In_2O_3$ is a semiconductor with the oxygen $p$-band filled. When a Cd-atom replaces a In-atom, the resulting system is metallic because of the lack of one electron necessary to fill the oxygen $p$-band. Comparison of Fig. 2.a, 2.b. and 2.c. shows that the presence of Cd produces the appearance of Cd-$d$ levels and impurity states at the Fermi level ($E_f$). The question is if the real system we want to describe provides the lacking electron or not, and if this point is relevant for the calculation. Here we will present results assuming that the system provides the lacking electron (hence, the impurity level at $E_f$ is in a charged state). To describe this situation we added one electron to the cell that we compensated with an homogeneous positive background to obtain a neutral cell.

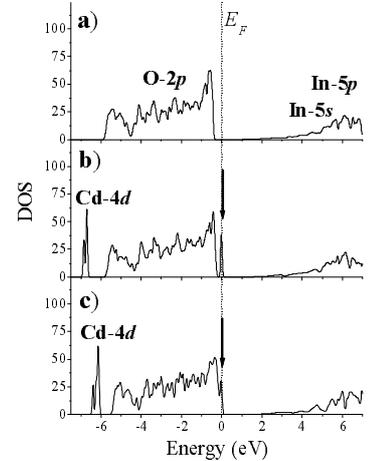

Figure 2: Density of states (DOS) for a) pure $In_2O_3$; b) One Cd impurity replacing a $D$–type In atom in $In_2O_3$. c) One Cd impurity replacing a $C$–type In atom in $In_2O_3$. In b) and c) the DOS correspond to the unrelaxed structures. Energies refer to the Fermi level ($E_F$). The arrows indicate impurity states in the valence band.

**Results and discussion**

Initially, we performed calculations in undoped $In_2O_3$. Since the $u$-, $x$-, $y$-, and $z$-parameters are free internal parameters, which turned out to strongly affect the EFG at both cationic sites, we refined the In($C$) and O positions by force minimization. The results ($u$=-0.0336, $x$=0.3910,

$y=0.1545$, $z=0.3813$) are in very good agreement with both X-ray diffraction results reported in the literature [9]. We also calculated the EFG tensor at both cationic sites in pure $In_2O_3$. As can be seen in Table 1, the obtained result for the EFGs at sites $C$ and $D$ are in relatively good agreement with those experimentally determined by PAC at Cd impurities [16]. This similarity can be understand from the similar electronic structure of Cd and In atoms.

After the study of undoped $In_2O_3$, we replace one $D$-type In atom (In($D$)) by a Cd. As can be seen in Table 1, the substitution of a indigenous In($D$) atom by a Cd impurity does not produce significant changes in the EFG. Again, this result can be understand from the fact that In and Cd present similar electronic structures. But, beside this similitude in the EFGs, the substitution of a In($D$) atom by a Cd impurity induce not negligible forces on the Cd-$O_{NN}$. In order to study the relaxation introduced by the impurity we considered the $O_{NN}$ displacements (assuming that structural relaxations preserve the point group symmetry of the cell in its initial configuration) until forces vanished. Displacement of atoms beyond the $O_{NN}$ are very small and none of the conclusions of this work are affected by it, thus we will not mention it in what follows. As can be seen in Table 1, the six Cd-$O_{NN}$ relax outwards along the Cd-O directions. The displacement of each of these O atoms is of about 0.1 Å (5% of the un-relaxed Cd-$O_{NN}$ distance). The magnitude of this relaxation was unexpected, and contradicts the assumption that Cd does not cause local lattice distortions in bixbyites [7, 8]. Concerning the EFG at Cd impurities located at cationic site $D$, the only effect of the relaxation is a small reduction of $V_{33}$. After relaxation, the results for the EFG are in excellent agreement with the PAC results [16]. We want to mention that the EFG orientation in Table 1 correspond to the case of Cd in the bixbyites $Er_2O_3$ and $Ho_2O_3$ [8]. Due to the similarities between the structures, we assume that the EFG orientation in $In_2O_3$ should be nearby the same that in $Er_2O_3$ and $Ho_2O_3$.

Table 1: Mayor components, $V_{33}$, of the EFG tensor at Cd site (in units of $10^{21}$ V/m$^2$), asymmetry parameter $\eta$ and $V_{33}$ orientation obtained in the FLAPW calculations (this work) compared with experimental PAC results at 300 K. $d$(Cd-$O_{NN}$) are the distances in Å from Cd to its $O_{NN}$. In the case of $In_2O_3$, $d$(Cd-$O_{NN}$) refers to the In-$O_{NN}$ distances. The experimental EFG tensor orientations correspond to $Er_2O_3$ and $Ho_2O_3$ (see text).

| Site $D$ | | | | |
|---|---|---|---|---|
| | $d$(Cd-$O_{NN}$) | $V_{33}$ | $\eta$ | $V_{33}$ orientation |
| $In_2O_3$ | 2.18 | +8.2 | 0.00 | [1;1;1] |
| $In_2O_3$:Cd unrelaxed structure | 2.18 | +7.8 | 0.00 | [1;1;1] |
| $In_2O_3$:Cd relaxed structure | 2.27 | +7.6 | 0.00 | [1;1;1] |
| Experimental [16] | - | 7.7$_1$ | 0.00$_5$ | [1;1;1] |
| Site $C$ | | | | | | |
| | $d$(Cd-$O_{NN1}$) | $d$(Cd-$O_{NN2}$) | $d$(Cd-$O_{NN3}$) | $V_{33}$ | $\eta$ | $V_{33}$ orientation |
| $In_2O_3$ | 2.13 | 2.19 | 2.23 | +5.4 | 0.98 | [0;-0.70;1] |
| $In_2O_3$:Cd unrelaxed structure | 2.13 | 2.19 | 2.23 | +4.8 | 0.96 | [0;-0.70;1] |
| $In_2O_3$:Cd relaxed structure | 2.20 | 2.29 | 2.33 | +5.6 | 0.68 | [0;-0.75;1] |
| Experimental [16] | - | - | - | 5.9$_1$ | 0.69$_1$ | [0;-1;1] |

We performed a similar study for the case of a Cd impurity located at cationic site $C$ of $In_2O_3$. Similar to the case of Cd at site $D$, the substitution of a In($C$) by a Cd atom does not produce significant changes in the EFG tensor (see Table 1), but the presence of the Cd impurity induce a strong relaxation of its six $O_{NN}$. The magnitude of this relaxation is different for each of the three pairs of distances Cd-$O_{NN}$, but in average is similar to those produced at the site $D$ (5%). As can be seen in Table 1, the structural relaxation changes the magnitude of $V_{33}$, in the symmetry of the EFG tensor and a small reorientation of $V_{33}$. The change in the symmetry of the EFG tensor reflects that the relaxation is non-isotropic. Finally, when the relaxed position of the Cd-$O_{NN}$ are taken into account, the agreement between our FLAPW results and the experimental ones [17] is excellent.

It is important to compare our results with those predicted by PCM. The results of these calculations are presented in Table 2. As can be seen, the agreement between PCM and FLAPW predictions is very good for the case of site $D$. From these results, it could be stated that the ionic model can be used for the description of the EFG at cationic sites in $In_2O_3$. Nevertheless, the PCM predicts a negative sign of the EFG for site $C$, in contradiction with the FLAPW results. If the relaxed positions obtained by FLAPW are introduced in the PCM, the discrepancy in the sign of the EFG at site $C$ is removed, but the agreement between the PCM, FLAPW predictions and experimental results for $V_{33}$ at both sites and the symmetry at site $C$ is not good. From this results, it is clear that the discrepancies between PCM and FLAPW (a method that take into account without the use of external parameters the electronic structure of the Cd impurity and the structural and electronic process that its presence induces in the $In_2O_3$ host) can not be associated only to the structural relaxations. We have to conclude that the purely ionic model in combination with lattice summations and the Sternheimer factor fails in the description of the EFG tensor at cationic sites in $In_2O_3$.

Table 2: $V_{33}$, $\eta$, and $V_{33}$ orientation obtained in our FLAPW calculations (relaxed structures), PCM, and PCM considering the structural relaxations predicted by FLAPW.

|  | Site $D$ | | | Site $C$ | | |
| --- | --- | --- | --- | --- | --- | --- |
|  | $V_{33}$ | $\eta$ | $V_{33}$ orientation | $V_{33}$ | $\eta$ | $V_{33}$ orientation |
| FLAPW | +7.6 | 0.00 | [1;1;1] | +5.6 | 0.68 | [0;-0.75;1] |
| MCP | +7.5 | 0.00 | [1;1;1] | -4.9 | 0.83 | [0;-0.75;1] |
| MCP + FLAPW relaxation | +6.7 | 0.00 | [1;1;1] | +4.6 | 0.55 | [0;-0.75;1] |

In conclusion, *ab initio* calculations using the FLAPW method successfully predict the experimental EFGs at Cd impurities in $In_2O_3$ and yielded quantitative information on the lattice relaxation around the impurity. From our results it is clear that the problem of the EFG at impurities in this semiconductor cannot be described by a point-charge model and antishielding factors. A proper theoretical description of electronic properties of Cd in bixbyites should consider self-consistently the electronic structure of Cd and the impurity-induced distortions in the hosts without external parameters or arbitrary suppositions. .

This work was partially supported by CONICET, Fund. Antorchas, ANPCyT (PICT98 03-03727), Argentina and TWAS (RGA97-057), Italy. L.E. and G.D. are fellows of CONICET. G.F. and M.R. are members of CONICET.